# Recent developments of selective laser processes for wearable devices

By *Youngchan Kim†, Eunseung Hwang†, Chang Kai, Kaichen Xu, Heng Pan\*, Sukjoon Hong\**

Y. Kim, E. Hwang, Prof. S. Hong
Optical Nanoprocessing Lab., Department of Mechanical Engineering, BK21 FOUR ERICA-ACE Center, Hanyang University, 55 Hanyangdaehak-ro, Sangnok-gu, Ansan, 15588 Korea

C. Kai, Prof. H. Pan
Multiscale Manufacturing Lab., J. Mike Walker '66 Department of Mechanical Engineering
Texas A&M University, College Station, TX 77843, USA

Prof. K. Xu
State Key Laboratory of Fluid Power and Mechatronic Systems, School of Mechanical Engineering
Zhejiang University, Hangzhou 310027, China

[\*]　　To whom correspondence should be addressed.
　　　E-mail: sukjoonhong@hanyang.ac.kr, hpan@tamu.edu

[†]　　Youngchan Kim and Eunseung Hwang contributed equally to this work.

Keywords: Selective laser process, wearable device, transformative approach, laser-induced graphene, ablation, sintering, synthesis

## Abstract

Recently, the growing interest in wearable technology for personal healthcare and smart VR/AR applications newly imposed a need for development of facile fabrication method. Regarding the issue, laser has long been proposing original answers to such challenging technological demands with its remote, sterile, rapid, and site-selective processing characteristics for arbitrary materials. In this review, recent developments in relevant laser processes are summarized in two separate categories. Firstly, transformative approaches represented by laser-induced graphene (LIG) are introduced. Apart from design optimization and alteration of native substrate, latest advancements in the transformative approach now enable not only more complex material compositions but also multilayer device configurations by simultaneous transformation of heterogeneous precursor or sequential addition of functional layers coupled with other electronic elements. Besides, more conventional laser techniques such as ablation, sintering and synthesis are still accessible for enhancing the functionality of the entire system through expansion of applicable materials and adoption of new mechanisms. Various wearable device components developed through the corresponding laser processes are then organized with emphasis on chemical/physical sensors and energy devices. At the same time, special attention is given to the applications utilizing multiple laser sources or multiple laser processes, which pave the way towards all-laser fabrication of wearable devices.

# 1. Introduction

The emergence of wearable devices can be attributed to several key factors and current global trends. Rapid advancements in microelectronics have made it possible to create smaller, faster, and energy-efficient components that can be integrated into a single device, while interest towards portable devices as well as new forms of electronics pushed the fabrication of electronics on flexible and stretchable substrates. From the perspective of consumer demands, people have shown a growing interest in health monitoring, and it is apparent that this trend will be continued in the future as the percentage of elderly people, who requires intensive and prolonged healthcare, is rapidly increasing in many countries. The outbreak of a pandemic, such as COVID-19, also brought a significant impact on the interest in and adoption of wearable devices.[1] These factors, among other keywords such as VR/AR, have contributed to the emergence and proliferation of wearable devices in various sectors, from consumer electronics to healthcare and beyond. The wearable technology industry is therefore expected to evolve continuously, with ongoing innovations and developments in response to changing consumer needs and technological capabilities.

Meanwhile, applying fabrication technologies developed for conventional electronics to new types of electronics, such as flexible, stretchable, or wearable devices, can pose several problems and limitations. The most apparent problem is that conventional electronics often involve high-temperature processes, which are not compatible with heat-sensitive flexible substrates used in wearable electronics.[2] These substrates may deform, melt, or degrade at elevated temperatures, making it challenging to apply traditional techniques. It should be also noted that functional materials[3] and novel structural designs[4] for wearable device are still under active investigation. The conventional fabrication technologies are often not fully compatible with new materials such as low-dimensional nanomaterials, and frequent design change that is inevitable at early developmental stage also acts as a burden for conventional photolithography-based fabrication methods. To address these challenges, researchers and engineers are developing new fabrication techniques tailored to the specific requirements of wearable electronics, and we believe that the selective laser process can make a significant contribution.

Laser is an attractive tool for material processing due to its unique properties as a light including monochromaticity, high energy density and tunability. As a result, laser-based materials processing already has undergone long-term investigation in scientific research,[5] whereas selective laser processing has been adopted in the prototyping of wearable devices recently, at a lab scale in particular.[6, 7] In this regard, we expect that it is timely to summarize current advances in selective laser process for wearable device, and would like to discuss the future directions of development. In this review, research on selective laser process for wearable device is divided into two sections – transformative, conventional – as shown in **Figure 1**. Since the discovery of laser-induced graphene (LIG) in 2014,[8] number of publications related to transformative approach has been increased at a fast pace. As a consequence, transformative approach has become an indispensable laser process for wearable devices, and its in-depth research not only expanded the applicable materials but also established new methods for creating more complex devices composed of multilayers. On the other hand, conventional laser process such as micromachining, ablation, texturing, sintering, synthesis and annealing has been also evolved constantly for fabrication of wearable device in line with transformative approaches. For more concentrated review on the latest studies, we mostly focus on the publications after 2020, and the ones require vacuum environment are omitted.

## 2. Transformative laser process

### 2.1. Design optimization

LIG has recently been in the spotlight in the field of wearable devices based on its excellent mechanical properties, electrical conductivity, three-dimensional (3D) porous structure, and low cost. LIG can be produced relatively easily through thermal decomposition and carbonization by irradiating a laser to a polymer material to cause a photothermal reaction.[8] LIG produced using a laser may have different shapes, chemical and physical properties depending on laser processing parameters such as laser power[8], working distance[9], hatch size, and spot size[10]. This is because the photothermal reaction is affected by the modulation of the energy density delivered by the laser to the material depending on the laser parameters. It is a well-known fact that the thickness of LIG increases as the laser power increases and that lower scanning speeds can produce better LIG with more uniformity and improved electrical conductivity. In this regard, the characteristics of LIG that vary depending on laser processing conditions enables multiple functions that are vital in wearable devices. In recent research, a triboelectric nanogenerator (TENG) and touch sensor were developed utilizing LIG in the form of long fibers (LF-LIG) generated through control of working distance of laser.[9] **(Figure 2a)** LF-LIG exhibits excellent triboelectric properties through a larger surface area and lower work function compared to conventional LIG. These properties made energy harvesting possible by attaching to the skin and producing electricity through simple contact. In addition, different functionalities can be obtained by changing the shape of LIG by simultaneously adjusting laser parameters such as spot size and hatch size.[10] As shown in **Figure 2b**, the arc-structured LIG shape can be obtained under high temperature and low-pressure conditions through a millimeter-sized laser spot, and at the same time, the wave-like array LIG is created by adjusting the hatch size. This structure exhibits characteristics that are sensitive to deformation due to stress concentration along a large radius of curvature. Using these properties, a foot pad sensor was developed that can monitor health status by producing a piezoresistive sensor.

Furthermore, the parameters of LIG can be altered not only through laser parameters mentioned above but also through laser patterning.[11] The selective processing capabilities of lasers enable complex design patterning and provide geometric performance through design optimization. For example, in **Figure 2c**, the LIG-based strain sensor fabricated under the same process conditions can be used as a sensor to detect wrist pulse since the sensor with the mesh pattern has relatively better sensitivity to resistance changes compared to the sensor without the pattern.[12] Alternatively, complex patterns, such as circles, also can be used to provide a temperature-sensing capability.[13] (**Figure 2d**) Using a circular pattern increases the surface area of the LIG and the resistance of the LIG, thereby further improving the temperature-sensitive response. Additionally, the patterning process can be applied not only to two-dimensional (2D) patterning but also 3D shapes[14], including one-dimensional (1D) fibers that are difficult to process[15, 16]. In the case of 1D LIG fiber, it can be assembled into numerous forms such as 2D nanosheets or plates based on its flexibility and thin diameter, while retaining excellent flexibility, rigidity, and lightness compared to other dimensions. These 1D LIG fibers can be utilized as multifunctional sensors for a variety of applications, including liquid detection, airflow, and respiration monitoring.[16] (**Figure 2e**)

In addition to the tunability of LIG mentioned above, it is possible to transfer LIG to other materials. The 3D porous structure of LIG allows transfer to an elastomer substrate while maintaining the shape of LIG through complete penetration of the liquid elastomer.[14] (**Figure 2f**) The LIG transferred in this way has not only secured its mechanical properties but also gained flexibility and elasticity, so it can be used under high deformation. LIG, which has such a 3D porous structure, is flexible and can be shaped into different forms, which is beneficial for wearable devices[17] or skin-attached sensors[18, 19] that require elasticity.

## 2.2. Different native material

In addition to the PI film[20], other cross-linked synthetic organic materials such as colorless polyimide(cPI)[21], polyamide-imide(PAI), and polyetherimide (PEI) are also capable of applying transformative laser process into graphene.[22] Beyond polymer materials, synthetic organic substances like Graphene Oxide (GO) can be transformed into graphene oxide (rGO)[23, 24] or laser-scribed Graphene (LSG)[25] through laser scribing processes, and utilized as flexible strain sensor[26] or flexible pressure sensors[27]. (**Figure 3a**) Although industrially synthesized precursor materials[28-31] as described above exhibit excellent mechanical and electrical properties, they often show poor biocompatibility and are difficult to reuse, making them less environmental friendly. To overcome these issues, researchers have recently shifted their focus towards developing LIG using easily obtainable natural materials such as paper[32], cork[33], and lignin[34], or natural fiber materials like cotton[35] and silk[36]. (**Figure 3b**) These natural materials offer advantages such as high carbon content, easy accessibility from natural sources, and reusability. As shown in **Figure 3c**, to transform natural materials into LIG, it is necessary to optimize laser specifications and process conditions by considering the photon energy absorption rate depending on the material. For instance, lignocellulosic materials such as wood, paper, and textiles tend to predominantly reflect near-infrared (NIR) light and strongly absorb ultraviolet (UV) light.[37] Due to these properties, using laser wavelengths in the UV region is more effective in LIG generation as it can induce a more efficient photothermal reaction.[37] Some heat-sensitive natural materials, such as silk and paper, can be carbonized through appropriate photothermal reactions to form graphene, but it can cause damage to materials once the laser condition is inappropriate. As a result, careful optimization is needed considering the safety of the materials.[36] Because natural material-based LIG is highly biocompatible and cost-effective, research is underway to develop functional devices that require biocompatibility and mass production, such as skin-attached patch sensors[38].

It is possible to produce sensors with various functions simply by making graphene using only natural materials, but even better performance can be achieved by combining natural materials with other synthetic materials[39]. Composites of natural and synthetic materials can be fabricated as follows: natural materials are mixed with polymer materials like PDMS[40] or aerosols[41], or with GO nanosheets[42]. This mixture can then be coated onto a substrate and solidified[43], or it can be made into a fibrous membrane using electrospinning[44]. Films or fibrous membranes produced through appropriate mixtures are easily converted to graphene by promoting a carbonization reaction caused by laser writing technology due to the natural material with high carbon content. When using natural materials alone, there are problems such as low mechanical strength, durability, or degradation from environmental factors such as moisture[32], but when mixed with synthetic materials, these problems can be overcome, which leads to improved performance with multifunctionality.

Additionally, researchers have been conducting investigation to produce LIG-based wearable devices using more advanced materials.[45] For example, they combined Kevlar and PVA/$H_3PO_4$ to create micro-supercapacitors[46] or developed a glucose sensor by electroplating copper on a Nomex insulating sheet induced to graphene *via* a laser[47]. (**Figure 3d**) In addition, research is being conducted on innovative LIG-based wearable devices using polyether ether ketone (PEEK)[48], an engineering plastic, and through this research, products such as wearable gloves capable of self-power generation and touch sensing are being developed. [49] (**Figure 3e**)

## 2.3. Simultaneous material transformation

In addition to the facile preparation of LIG from various organic materials as mentioned above, the composite configuration of LIG with certain nanomaterials is also readily fabricated by introducing mixed organic precursors for simultaneous material transformation.[50, 51] The composite precursors can be either prepared by blending nanomaterials with polymer bases such as poly (amic acid) (PAA)[52] and polybenzimidazole (PBI)[53] before solidification or directly coated on the substrate[54, 55]. Although the preprocessing steps show minor differences, the precursors eventually form planar mixture films, through coating and baking procedures, to obtain LIG/nanomaterial composites with a single scan of a focused laser. (**Figure 4a**) Compared to the standalone use of LIG, the resultant nanocomposite accepts new materials to significantly advance the inherent performance[56] or provide unique characteristics[57] exceeding the innate versatile functionality of the LIG[58]. The reported nanocomposites generally consist of a heterogeneous structure embedded with low-dimensional materials (LDM), which are highly recognized for excellent properties[59] such as large surface area, ultralight weight, high reactivity to surrounding environments, and superior mechanical strength with flexibility. Regardless of their morphological diversity, all types of LDM are compatible with selective laser processes for simultaneous material transformation, including metal and metal oxide[60-62] nanoparticles in 0D, carbon nanotubes (CNT)[63] in 1D, and MXene nanosheets[64] in 2D. (**Figure 4b**) The above examples commonly utilize a single laser source to produce the composite structures of conductive LIG channels with sensitive LDM nanomaterials for mechanical/chemical sensors and energy devices.

While the laser-induced material transformation is primarily a thermal process, however, the underlying heating mechanisms depend on the laser output operation mode, represented by continuous-wave (CW) and pulsed laser. Currently, the ultrashort pulsed laser utilizes the femtosecond (fs) repetition rate in pulse modulation which minimizes the heat transfer to the surroundings during the laser irradiation.[65] Upon ultrafast photonic dynamics, when free electrons transfer energy to lattice by electron-phonon collision[66] at single-pulse irradiation, the target accumulates generated heat through the multiphoton absorption effect from fs laser pulses.[67] On the other hand, the successive beam propagation of a CW laser generates a distributed temperature field, considered a heat-affected zone, by the photothermal effect extended to the periphery of the irradiated spot through the material heat transfer.[68] In this connection, the difference in photo-thermal-chemical reaction of each laser operation mode presents distinct micromorphology in resulting LIG/nanomaterial composite construction. As shown in **Figure 4c**, the fs laser-processed area possesses a uniform graphene structure with ZnO particles remaining on the surface while the CW laser-processed area holds irregular microstructures with aggregated ZnO clusters intensively mixed in carbon molecular skeleton. As a result, relatively smooth LIG surfaces, capable of NIR light detection by using the bolometric effect of graphene[69], and the embedded ZnO, utilized for UV light detection through the photoconductive effect[70], realize successful implementation of the dual-mode photodetector.[71] (**Figure 4d**) It is noteworthy that the introduction of multiple laser sources has proven its ability to effectively tune the function of the device, which will be further discussed for advanced selective laser processes toward all-laser technology.

For wearable applications, on the other hand, mixing PI with stretchable organic substrates plays an important role in adopting the intrinsic flexibility of the substrate to the brittle carbon products by completely integrating the LIG into the substrate through concurrent carbonization. One of the most promising substrates for LIG integration is PDMS since the exceptional transparency, stretchability, and biocompatibility are secured at once.[72] The PI/PDMS composite with specifically designed LIG electrodes shows adequate electrical properties and durable mechanical robustness for skin-attachable strain sensors.[73, 74] Furthermore, the paper- and cloth-integrated LIG circuit[75] allows great potential as functional building blocks such as packaging components and fabric fibers for smart wearable electronics applied in our daily lives. In conclusion, we expect that the laser-induced simultaneous material transformation technique will alleviate the labor-intensive works of the conventional LIG-based wearable device fabrication methods by the simple assembly of LIG with novel nanomaterials and favorable substrates. Yet the practical skin-interfaced devices at a systematic level inevitably require additional processing steps to create sophisticated electronic circuit designs to achieve desired objectives.

## 2.4. Consecutive multilayer configuration

Based on the precisely patterned LIG layers, functional multilayers are assembled by consecutively stacking on-demand materials to meet the requirements for the successful operation of the device in corresponding circumstances. In the simplest cases, the LIG patterns work as a conductive electrode or multifunctional element and the applied functional materials are responsive to intended stimulations while the encapsulation layers are selectively added to protect the entire system if needed. The representative applications are organized as energy devices, environmental sensors, and health monitoring systems with selected materials as listed below: 1) Supercapacitors using PEDOT[76] or phosphor Cu[77], biofuel cells[78] using glucose dehydrogenase bioanode and bilirubin oxidase biocathode, and TENG using GO[79] or pressure sensor integrated TENG using rGO[80]. 2) Mechanical sensor using cobalt nanoparticles[81] or Ag electrodes with kirigami design[82], humidity sensors using GO[83] or $Pd/HNb_3O_8$[84], and gas sensors using $MoS_2$[85] or moisture-resistant gas sensor using PDMS[86] as a water molecules repellent and semipermeable membrane layer. 3) Integrated skin-interfaced monitoring system[87] consists of $ZnIn_2S_4$ (ZIS) for skin moisture sensing with $CNT/SnO_2$ composite for simultaneous temperature sensing and sweat sensors using polymers[88, 89] or metals[90-95]. The above-listed research equally highlights the flexibility and stretchability of the demonstrated device as a proof-of-concept for wearable applications and displays the guaranteed scalability of the LIG-based multilayer configuration toward multiple conjugations of extra components including electronic and mechanical systems.

The advanced form of attachable sensing devices features highly sophisticated multilayer circuit configurations for multimodal sensor networks and wireless communication. As shown in **Figure 5a**, a flexible plant growth monitoring system is comprised of a temperature sensor, an optical sensor, and two humidity sensors for real-time measurement of the environmental temperature, light irradiation intensity, and humidity.[96] The screen-printed Ag and LIG electrodes functionalize each sensor, combined with ZIS nanosheets or $SnO_2/CNT$ film, and arrange conductive channels to interconnect electronic elements. Due to the cellular structures of the porous LIG electrodes depicted in the detailed schematic of the integrated device, active humidity sensing of ambient humidity and leaf transpiration are monitored owing to the absorption and desorption of the water molecules. (**Figure 5b**) One another advantage of the fully integrated sensing chip system is the improved applicability in versatile multi-scenario situations. The Ag/MXene sponge incorporated in the LIG-based interdigitated electrode realizes a fully integrated wireless pressure sensing chip system by comprising a flexible printed circuit board composed of various electronic chip components including capacitors, resistors, and data transmission antenna.[97] (**Figure 5c**) **Figure 5d** illustrates the applications of the proposed attachable pressure sensor in different scenarios such as fruit growth, joint movement, and pulse signal thus promising in the next generation wearable devices.

Beyond the progress of functional electronic circuit designs, the implantation of individual rigid systems on the wearable scaffolds, implemented by virtue of selective patterning of LIG on flexible PI films, allows a high level of biomedical signal analysis. As a specific example, the heterogeneous combination of narrowband vertical cavity surface emitting lasers (VCSELs) and a Si photodetector (PD) unit on the flexible Au-LIG hybrid electrode constitute epidermal biocompatible optoelectronic sensor for health monitoring.[98] (**Figure 5e**) On the epidermis surface, VCSELs emit lasers in separate wavelengths that reflect back to the PD after passing through the blood vessel which measures the health information of the wearer such as blood oxygen level and pulse rate. (**Figure 5f**) Moreover, to offer reliable medical diagnostics by investigating human sweat, compositive data containing multiple chemicals in human sweat are required while the conventional wearable sweat sensors could not provide valuable detection since the studies were focused on proof-of-concept designs. For the scalable and accurate use of the sweat sensor, a microfluidic chamber installed to supply consistent and uncontaminated sweat volumes to the LIG-based sweat sensor enables continuous multisubject analysis of both sweat metabolites and electrolytes.[99] (**Figure 5g**) In **Figure 5h**, the real-time response of on-body glucose, lactate, and sodium are investigated during cycling to monitor the physical state. The announced reports well-emphasize the necessity of consecutive multilayer configuration on basic LIG electronics for feasible applications of wearable devices by introducing functional materials, advanced circuit designs, and prepared adaptable systems.

## 3. Conventional laser process

### 3.1. Laser ablation/machining

Selective removal of materials using laser has distinct advantages over other standard fabrication techniques especially for unconventional materials[5] with complex compositions or other emerging materials that requires rapid prototyping for proof-of-concept wearable devices. Laser ablation technique therefore has been effective patterning scheme for low-dimensional nanomaterials together with their nanocomposites which have been extensively investigated as constituents of flexible and stretchable applications. We have witnessed that recently introduced nanomaterials and nanocomposites including Cu-Au core-shell NW,[100] Ti3C2−MXene,[101] PDDA modified reduced graphene oxide with exfoliated Ti3C2Tx,[102] PEDOT:PSS with aramid nanofibers,[103] MWCNT/PDMS composite[104] and MWCNTs-MnO2[105] film are now compatible to laser patterning techniques based on the conventional ablation mechanisms. Majority of these new materials can be removed at microscale resolution with clean features and minimized damage to the underlying substrate once laser parameters are optimized as shown in **Figure 6a**. The thickness subject to the ablation is in the order of ~100 nm in the case of NW percolation network,[100] yet the laser ablation is compatible to considerably thicker target layer even up to several hundreds of micrometers.[102] The as-prepared patterned layer often undergoes additional processing such as electroplating,[105] transfer to other thermal plastic film,[102] or direct attachment to the human body[100] in order to be applied as electrophysiolgial sensors for human motion monitoring or energy sources

for stand-alone operation of other wearable devices. The sensor can be divided into

Facile patterning capability enabled by laser ablation allows rapid design change, which also brings changes in device performance as shown in the case of microsupercapacitor with the electrodes at parallel and series connections for different working voltage.[103] It should be noted that the alteration of working voltage also can be achieved by the application of laser process at the electrolyte layer instead of modifying the electrode design.[50] Apart from the patterning capability, laser ablation has been advanced in multiple directions: laser ablation itself can affect the performance of the resultant device even without design change as confirmed from the study on laser ablation of MWCNT/PDMS composite.[104] It is presented that the cross-section morphology of the laser-ablated sample and hence the resultant conductive network is adjusted by the laser condition used for the ablation process, which is also closely related to the gauge factor of the final strain sensor. (**Figure 6b**) In-depth study on other specific ablated targets such as graphene films of 6-8 layers[106] also shows that different ablation patterns may bring distinct working mechanisms for each usage. The applicable material is continuously expanding at the same time, which now includes uncommon microbial conductive biofilm such as *Geobacter sulfurrenducens* for electricity generation from water evaporation.[107] From the perspective of applications, thin-film thermoelectric (TE) generator and stretchable battery have been newly developed by patterning TE materials in zigzag pattern[108] and separating lithium titanium oxide (LTO) anode/lithium iron phosphate (LFP) cathode into independent microscale square array[109] (**Figure 6c**) based on direct laser ablation technique. These studies in turn successfully broaden the range of laser-enabled wearable energy devices, which has been mostly focused on microsupercapacitor to date.

Although laser ablation is increasingly applied to new materials as mentioned above, the mechanism behind laser ablation remains analogous, involving diverse interactions and feedbacks between incident light and the target material.[5] On the contrary, recent works on laser micromachining of specific substrates describe attractive two-step approaches that incorporate transformative step before material removal. It remains unclear which class of substrates are compatible with the corresponding approach, yet meaningful results have been found in PDMS and PI, which are both valuable materials for wearable device applications. It should be noted that conventional laser ablation techniques are still extensively utilized for the relevant materials[110] in order to create various on-demand patterns including quasi-3D patterns[111-113] and rapid prototypes of user-defined microfluidics channels.[114, 115] The conventional laser ablation using pulsed laser, however, yields limited surface quality with residual burrs in general even with ultrashort laser[116] due to uncontrollable ablation phenomena. On the other hand, recently developed patterning technique on PDMS[117] takes advantage of seamless photothermal pyrolysis conducted by continuous-wave (CW) laser, which converts PDMS into SiC along the laser scanning path. Although PDMS is highly transparent at visible wavelength, 532 nm green laser is utilized to induce selective pyrolysis by the aid of successive laser pyrolysis (SLP) mechanism that relies on the iterative change in laser absorption and subsequent increase in heat transfer to the vicinity upon the conversion of PDMS into SiC. The resultant SiC is then simply removed by mechanical means such as ultrasonication to complete non-ablative PDMS machining process as shown in **Figure 6d**. The resultant PDMS after the machining is not shows exceptional surface morphology, (**Figure 6e**) but also confirmed to possess its original surface chemistry, which is essential for customizable organ-on-a-chip applications. Non-ablative micromachining technique for PI[118] substrate is equivalent to the case of PDMS, but two major differences are observed: The product created by laser-induced pyrolysis is LIG for PI, and it can be exfoliated spontaneously from the original matrix due to the intensive gas emission at the laser spot under scanning. This 'pyrolytic jetting' technique not only enables direct micromachining of PI substrate, but also can be employed to create 3D LIG helical spring[119] (**Figure 6f**) through a minor experimental modification, which is expected to be useful for multifunctional electromechanical systems as confirmed from its strain-force characteristics.

From a practical point of view, laser equipment has become more widely available and emerged as an efficient tool to create proof-of-concept wearable device promptly. As a result, the use of laser machining keep increasing even for conventional materials. For instance, laser ablation of Cu thin film in specific patterns, e.g. serpentine, enables quick feasibility tests of stretchable wearable applications such as thermos-haptic device[120] and multispectral active cloaking device.[121] We also can find indirect usage of Cu as a mold after laser machining into microcone cavities for sensitivity tuned pressure sensor.[122] Along with these attempts, laser process is actively integrated with different forms of industrialized material for wearable applications such as denim fabric[123] and textiles[124] for automated, data-driven mass production of real-world wearable devices. Meanwhile, theoretical and experimental studies to develop novel method for laser selective removal of multilayer materials are also actively conducted[125] at the same time.

### 3.2. Laser texturing

Laser texturing can be regarded as one of the subsections of laser machining process, yet laser texturing is more focused on the modification of the surface properties in general. In particular, laser texturing is a facile and effective method to change the wettability of the target material,[126] and recent studies show that the laser-induced wettability control enables direct patterning of liquid metal without injecting the fluid into a predefined microchannel, which is the most widely used method for EGaIn based stretchable devices. Once femtosecond laser microfabrication is applied to a PVA substrate, super-metal-phobic area based on porous network of micro/nanostructure is created[127, 128] while the functional groups of PVA remains unchanged. Due to the modified surface morphology, the oxide layer of EGaIn and PVA, which originally is at Young's state, changes into Cassie state at the contact angle of 162.5° as shown in **Figure 6g**. As a consequence, EGaIn is only printed at the pristine PVA without laser texturing upon brushing. An additional advantage of the corresponding EGaIn electronics is that the circuit can be repaired or even reconfigured *via* rebrushing of EGaIn droplet. Wettability control by laser texturing also can assign new functionality or enhance the performance of final device as shown in the case of texturing double-side micropyramid on PDMS by laser.[129] Meanwhile, laser-induced surface texturing allows patterning of functional layer even without changing wettability by using PE film as supplementary layer to create groove at the target substrate.[130] Laser texturing is also compatible to other laser process: by applying laser texturing on LIG, non-stick conductive structure for Galinstan droplet is created to develop a tilt sensor that can distinguish various slanting orientations.[131]

Surface texturing often is crucial for some applications, and a representative example is surface-enhanced Raman spectroscopy (SERS) that allows high-precision detection of target molecules.[132, 133] SERS techniques are expected to be vital for wearable healthcare monitoring devices and sensors for real-time diagnosis and detection of potential threats. Recent studies show that laser texturing enables successful fabrication of efficient SERS template. Since a sensitive SERS template should be able to acquire signal from a highly diluted solution, it is crucial to suppress coffee ring effect during evaporation by controlling contact angle (CA) and sliding angle (SA) at appropriate ranges, which is realized by femtosecond laser texturing of AISI304 stainless steel sheet.[134] A more advanced version of wearable SERS template is created on 0.1 mm Cu foil by creating superhydrophobic/superhydrophilic hybrid structures with micropore at the center by laser in order to deposit target material precisely at the designated area as shown in **Figure 6h**.[135] As a result, the resultant SERS template is capable of detecting target molecules even at attomolar level, proving high potential of selective laser texturing techniques for highly sensitive biosensors.

### 3.3. Sintering and synthesis

As it is also mentioned in laser ablation/machining section, nanomaterial is an important class of materials for wearable devices. Laser technologies such as sintering and synthesis have been investigated for efficient use of these materials in additive manner.[7] Due to large surface area exhibited by nanomaterials, noble metals – Ag NP and Ag NW in particular – have been the center of research, and a large portion of recent research can be regarded as an extension of this trend,[136, 137] supported by their own originality. Most of laser processing to date has been applied to a flat surface even for flexible and stretchable applications, yet Ag NP laser sintering is recently conducted on diverse types of substrates such as PVDF monofilament fiber with circular cross-section.[138] It is even possible to separate donor and acceptor substrate to expand the applicable polymer substrate for laser sintering process.[139] Laser sintering also becomes more compatible to more advanced applications such as data-driven motion sensor and multifunctional skin electronics.[140, 141] Along with this overall development status, some selected publications that indicate two different development perspectives are introduced: modification of the sintering process and introduction of additional emerging material.

Laser sintering is now applied to wide range of metal NPs, while analogous reductive sintering process also has been reported for several metal-oxide NPs such as CuO and NiO NP, allowing facile generation of electrodes at microscale. In this connection, a monolithic laser reductive sintering (m-LRS) is proposed through a minor change added to the reductive sintering process of NiO NP.[142] To create a planar electrode, hatch-scanning is incorporated in general to fill the designated area. In this study, several scanning lines are skipped intentionally during the hatch-scanning procedure to create NiO channel between two Ni electrodes as shown in **Figure 7a**. As a result, Ni-NiO-Ni structure is created to show negative temperature coefficient (NTC) thermistor, (**Figure 7b**) which is distinguishable from temperature dependency of Ni electrode that exhibits positive temperature coefficient of resistance. Interestingly, the B-value of the resultant Ni-NiO-Ni structure is calculated to be 7,350 K, and this value even reaches 8,162 K near room temperature. These values are known to be one of the highest sensitivity for thermistor-based temperature sensors, and it is claimed that such superior sensitivity might be originated from rapid thermal annealing experienced at the NiO channel that results in decreased nickel vacancies. Demonstration of the corresponding Ni-NiO-Ni as an attachable epidermal temperature sensor potentiate that m-LRS process can contribute greatly to the development of ultrathin wearable flexible devices.

In terms of relevant material, EGaIn is receiving special attention for laser sintering. Liquid metal is firstly prepared in the form of discrete droplet to be compatible with laser sintering scheme.[143] Liquid metal in droplet allows direct patterning of the liquid metal onto the target substrate, however, Ag NW is introduced in addition for enhanced adhesion between target material and to gain tunable electrical properties.[143] EGaIn particles (EGaInP) with Ag NW additive (AgNWs-EGaInPs) forms a stable film when the vacuum filtration and transfer method is applied, while EGaInP only results in a poor-quality film. The as-prepared Ag-NWs-EGaInPs film is composed of EGaInPs and AgNWs that are physically mixed, but their level of entanglement can be controlled by the subsequent laser irradiation. This entanglement structure allows not only strain-insensitive electrode that is crucial for wiring purpose, but also freestanding patterned liquid metal thin-film conductor.[144] (FS-GaIn) It is further confirmed that the freestanding liquid metal-AgNW composite thin film can be useful as an electrical connection between the chip and the conductor[145] that has remained as a major limitation for wearable devices. Once the pre-patterned freestanding stretchable solder sticker (STicker) is soldered between two conductors, the corresponding Sticker can be elongated even up to 5 mm as shown in **Figure 7c**.

Based on these developments, laser sintering is constantly becoming a mature technology that is now compatible with wearable devices at a higher complexity. One representative example shows this tendency is Ag NW and TLC-based Artificial Chameleon Skin (ATACS)[146] developed recently for rapid coloration of the skin into the underlying habitat as shown in **Figure 7d**. Closer look into its structure reveals that the device is composed of dense multilayers, which are connected *via* laser sintering of Ag NPs. Meanwhile, for more site-specific, locally confined use of nanomaterial, laser-induced synthesis has been investigated for multiple NWs on flexible substrate.[147] For wearable device applications, two different types of NWs, i.e. p-CuO NW and n-ZnO NW, are consecutively synthesized by laser-induced hydrothermal growth scheme to create highly efficient UV micro-photodetector compared to the homojunction case.[148] While on the other hand, continuous efforts are made to synthesize new target materials such as SiC microwire[149] and Pd-$WO_3 \cdot xH_2O$ microwire.[150]

### 3.4. Others

Laser is a flexible tool that brings wide range of photothermal, photochemical and photophysical reactions. As consequences, the use of laser is continuously increasing in recent literature. Some latest examples that are relevant to wearable devices include IZO annealing,[151] curing of water soluble polymer,[152] sol-gel transition,[153] metal textile welding,[154] cPI lift-off,[155] and PR patterning.[156] In many research, laser simply provides facile substitutional treatment for conventional method, while it is also possible to bring improved performance compared to the one fabricated by other processes. For instance, **Figure 8a** shows the photocurrent density of a hematite nanorod (NR) photoelectrochemical cells as a function of applied potential after experienced different annealing processes. Apparently, thermally annealed hematite NR (THAN) shows inferior performance compared to the laser-annealed hematite NR (LAHN) even at high annealing temperature of 800 °C. This study proves that laser technique is not only an alternative process to the conventional counterpart, but also possesses high potential to bring significant improvements in the resultant device performance.

Meanwhile, it should be noted that innovative laser process is constantly reported, and some selected works relevant to wearable device are summarized in this section. One specific form of material that is valuable for wearable device is fiber for fabric and textiles, and thermal drawing technique (**Figure 8b**) allows integration of multiple elements such as tin selenide (SnSe) thermoelectric material into an arbitrary fiber. On the contrary, the performance of as-prepared thermoelectric fiber as a thermoelectric device is not satisfactory due to its polycrystalline nature. In this connection, laser is an appropriate tool to induce continuous crystallization of the thermal-drawn fiber into single-crystal SnSe fiber.[157] Laser crystallization successfully improved the ZT value up to 2 as shown in **Figure 8c**, which is comparable to the best ZT for bulk thermoelectric materials. At the same time, it is observable that the mechanical properties of SnSe fibers are compatible with the wearable device applications. Likewise, laser photothermal reaction can create distinct result according to the material under concern, often accompanied by additional process or specific configuration. For Cu NW, laser oxidation/reduction is combined with wet oxidation to complete reversible Selective Laser Induced Redox (rSLIR) cycle, which successfully creates multispectral monolithic photodetector that shows distinct response towards RGB signal. (**Figure 8d**) Regarding specific configuration, laser-induced photothermal reaction at the PI/PDMS interface reveals that adhesive-free bonding may occur due to the enhanced mechanical interlocking at microscale.[158] Laser-irradiation at PEDOT:PSS,[159] on the other hand, discovers that phase separation occurs to separate connected PEDOT-rich domain and PSS-rich domain as shown in **Figure 8e**, obtaining highly conductive water-stable hydrogel as a result. These latest studies suggest that laser processing is not only useful for wearable device, but also has the potential to bring innovations in bio implantable devices.

## 4. Applications

In the previous section, we have organized the latest development of selective laser processes in wearable technologies by focusing on the patterning of the multifunctional LIG with compatible nanomaterials, integration of sophisticated electromechanical systems to fabricate advanced functional devices, and employment of conventional laser techniques for feasible industrial applications. A plethora of applications intended for wearable device have been realized - at least to the proof-of-concept level - through the corresponding laser processes, and two most important categories of these laser-enabled wearable applications are sensors and energy devices, indicating that a primary goal for wearable device is to detect human motion and physiological data in a remote, standalone manner. The latest research on these sensors and energy devices are briefly summarized in terms of the relevant laser process, utilized laser source, materials under concern and resultant performance as the tables. Sensors can be subcategorized into chemical sensors, (**Table 1**) which analyze components such as body fluids or monitor the surrounding environment for gases and humidity, and physical sensors that respond to other factors such as strain and pressure. (**Table 2**) The energy devices includes energy harvesting devices including thermoelectric devices and triboelectric nanogenerators as well as energy storage devices such as supercapacitors and batteries. (**Table 3**)

While on the other hand, the emerging all-laser fabrication strategies, being considered as an ultimate goal of the selective laser processes by encompassing the broad advantages of individual laser-based methods, are of great interest for their noticeable impact on systematic laser processing in wearable device configurations. In this section, several advanced system-level research will be categorized into two representative aspects which are the multistep laser processing by sequential combinations of laser processes and the multi-laser processing by utilizing separate laser sources.

## 4.1. Multiple laser processes

Without complicated mechanical machining and restrictive chemical treatments, the all-laser-driven fabrication method successfully delivers productive outcomes available to alternate time-consuming procedures and toxic chemicals. The sequential fabrication of an on-body sweat sensor, resistive temperature sensor, and piezoresistive strain sensor by the laser engraving of LIG and microfluidic channels.[160] The excellent electrochemical sensing properties of 3D porous LIG structure allow real-time detection of uric acid (UA) and tyrosine (Tyr) in sweat which are important diagnostic indicators for disease control of gout and other metabolic disorders.[161, 162] While the precise investigation of sweat UA and Tyr is considered a big challenge since their contents are extremely low, however, the accelerated electron transfer[163] and catalytic activation for enhanced redox reactions[164] through an inevitable presence of abundant defects make LIG an ideal candidate for electrochemical biosensors. Supported by the fully integrated microfluidic chamber system, continuous operations of contamination-free collection and supply of sweat at a constant rate and chemical sensing are well-exhibited. For the vital signal monitoring, the strain-resistive LIG temperature sensor outputs an improved conductivity as the temperature rises owing to increased electron-phonon scattering speed and thermal velocity of electrons[22] while the 3D porous strain sensor simultaneously reads the external strain deformations. Thus, the multistep all-laser processing enables wearable multimodal biosensors with efficient microfluidic system integration for sensitive sensing under stable sweat sampling and multiplexed vital sign monitoring.

Going one step further, the additional laser processing may modify the original surface state and determine the material characteristics to the programmable level by controlling the laser operating parameters. **Figure 9a** clearly shows the distinct differences in surface wetting conditions between as-transferred LIG and the laser-textured LIG that the produced air gaps on the textured LIG surface support the Galinstan droplet to generate a Cassie-Baxter wetting state.[131] According to the relationship between the laser power and the hydrophobicity of the laser-textured surface, the higher the laser fluence is the deeper the texturing trenches are formed which increases both the contact angle and the sliding angle that eventually realizes the free-rolling state of the Galinstan droplet without the residue pinning on the LIG surface. In this regard, the fabricated tilt sensor reliably detects the human body motion upon repetitive random movements when the multimodal circuit configuration is completed by integrating the LIG/ZIS humidity sensor, LIG/PDMS strain sensor, and wireless communication system on a thin PI film. (**Figure 9b**)

Meanwhile, another useful combination of sequential laser processes is the UV laser ablation of cPI for stretchable designs and the annealing of metal nanoparticles by laser sintering for conductive electronic channel fabrication.[140] (**Figure 9c**) The 2D thin serpentine-patterned skin-interfaced sensor generates initial cracks on the laser-annealed layer before being attached to the skin, for the activation of the entire sensory system. After attaching the sensor covering the wide range of the wrist, the relative resistance responses against the epicentral deformation from the complex strain stress of five fingers are decoded by the deep-learning algorithm to provide real-time human motion data. (**Figure 9d**) For the fine-tuning of the detection sensitivity, the laser fluence can modulate the porosity of sintered nanoparticle structures by adjusting the annealing degree and thus the programmable crack propagation depth determines the gauge factor of the sensor. (**Figure 9e**) As a consequence, the detection of minute skin deformation of epicentral reaction from remote body sites is readily measured in low laser fluence conditions by the suitable laser-induced crack-based strain sensor, collaborated with deep-learning technology. In summary, the multistep processing of the all-laser fabrication methods not only reveals their possibility as a facile alternative for inefficient procedures but also substantiate their outstanding performances and biocompatibility for skin-interfaced wearable applications.

### 4.2. Multiple laser sources

Recently, researchers proposed an innovative wearable device processing technology that enables the development and optimization of thin-film electronic devices using two lasers with different wavelengths.[141] The proposed technology is based on two processes of multi-laser processing: (**Figure 10a**) 1) additive manufacturing through visible laser sintering of metal nanoparticles(NPs). The 532 nm continuous laser irradiates the AgNPs coated on the CPI film, selectively sintering the AgNPs into a designed metal pattern and patterning them into an electrically connected layer. 2) Cutting processing through ultraviolet laser ablation. The top AgNP electrode is selectively removed without damaging the substrate and the previously patterned bottom layer electrode via a 355 nm pulsed laser. This is based on the fact that materials absorb and transmit light differently depending on its wavelength. AgNPs show a high absorption rate of light with a wavelength of 532 nm and are sintered through a photothermal reaction. On the other hand, the low-power 355 nm laser can ablate AgNP electrodes with strong peak power but does not penetrate the layer and does not damage the substrate. Because electrodes can be easily created and deleted by these multi-lasers, they can modify the electrode patterns into the desired pattern at any time during the fabrication process. The attachable wireless networking multimode wearable device developed through this process can optimize impedance depending on the operating frequency. As shown in **Figure 10b**, impedance, which varies depending on the attachment location and user, can be optimized for the operating frequency by modifying the design of the antenna manufactured through the multi-laser process in real-time. In addition, the researchers developed wireless wearable devices that operate without batteries through RF energy harvesting through impedance optimization, demonstrating the possibility of commercializing the development of equipment such as VR and skin devices. (**Figure 10c**)

Another recent study developed a completely soft self-driven vibration sensor (SSVS) through a multi-laser process.[165] As shown in **Figure 10d**, SSVS is produced by combining PDMS and PI films through various processes such as laser pyrolysis, laser engraving, and laser texturing. The manufacturing process of SSVS is as follows. First, to collect electrical signals resulting from the vibration of Galinstan droplet, LIG electrodes patterned by photothermal reaction were created on PI film using an infrared laser. The LIG electrode is transferred to flexible and stretchable PDMS. LIG electrodes transferred to PDMS are then textured using a low-thermal effect UV-pulsed laser to form a non-wetting structure. Additionally, to protect the Galinstan droplet, a pulsed laser is also used on the PDMS cap to create a pattern through engraving. This is related to the fact that pulsed lasers do not carbonize LIG and PDMS but simply ablate them. LIG electrodes exhibit an electrical insulation effect and have high electrical resistance due to the destruction of the electrodes by the high energy intensity of the UV-pulsed laser. The texture of PDMS affects the contact angle with the Galinstan droplets. These properties can be further fine-tuned depending on the parameters of the laser to further improve the functionality of the sensor. As the laser fluence increases, the notch of the LIG electrode becomes deeper, which increases the specific surface area. This results in a signal amplification effect depending on the friction charging efficiency with the Galinstan droplet. Additionally, dense, deeply textured surfaces can be achieved by controlling the hatch size and laser fluence, which helps the Galinstan droplet maintain a stable shape with the surface while maintaining a large contact angle. Sensing fine vibrations through stabilized Galinstan droplet, the Surface Vibration Sensing System (SSVS) is integrated with the Bluetooth module to detect human movements through an application that monitors subtle vibrations of liquids in real-time and is equipped with a system that identifies emergency situations such as falls and slips. (**Figure 10e**)

## 5. Conclusion

In this review, selective laser processes that are applicable for the fabrication of wearable devices are classified and summarized to assess the status of the relevant technologies. For the transformative approach represented by LIG, in-depth studies are conducted in terms of design optimization and material selection, while use of additional materials allows rapid generation of more complex, hierarchical nanostructures that brings enhanced performance or additional functionalities to the device. In the case of conventional laser processing such as micromachining, sintering and synthesis, notable improvements are found *via* the expansion of applicable materials, smart modification of the existing method and discovery of novel applications enabled by the corresponding schemes. The listed laser processes act as core fabrication methods for individual elements that are vital for wearable applications including human motion detection, energy harvesting, sweat sensor and real-time health monitoring, to mention only a few. As the use of laser for fabrication of wearable device is continuously increasing, we expect that the selective laser process will be indispensable part for ambient manufacturing of multifunctional wearable devices in the future.

From the latest wearable devices mentioned in the application section, future development directions required for the laser process can be anticipated. We firstly witness that the recent studies on wearable devices not only demonstrate the proof-of-concept functionalities of individual components but also move towards the integration at the system level, which is much more challenging. For instance, to accomplish stand-alone operation of wearable sensor device at the system level, data acquired from a sensor has to be processed and transmitted, which requires a number of additional parts besides a sensing element such as microcontroller and converters.[160] As a consequence, compatibility of laser process with other fabrication techniques as well as intimate electrical contacts between distinct elements created or placed by different processes should be taken into account. It is also mentioned that more than two laser processes are employed in the selected studies, which suggest that all-laser fabricated wearable device is an attractive concept.

On the other hand, it appears that the full potential of laser processing has not yet been realized. Currently, laser processes used in the production of wearable devices largely rely on accessible lasers without detailed investigations on the laser conditions such as wavelength, pulse width and repetition rate. However, it is well known that the characteristics and hence the performance of the resultant device vary significantly depending on the laser conditions even for the same material.[166-168] In this regard, we expect that research institutions with specialty in laser should more actively involved in the relevant projects. High power laser experts are also essential in order to find a solution to the persistent issue of low throughput in laser processes. Through these research activities and professionals, we expect that the development of laser processes for wearable devices and its production can take a significant step forward.

## Acknowledgements

Y. Kim and E. Hwang contributed equally to this work. This work was supported by the Basic Research Program through the National Research Foundation of Korea (NRF) (No. 2022R1C1C1006593 & 2022R1A4A3031263). This work is partially supported by the National Science Foundation under Grant No. 2054098 and Grant No. 2213693.

## Data Availability Statement

Research data are not shared.

## Conflict of Interest

The authors declare no conflict of interest.

**Figure**

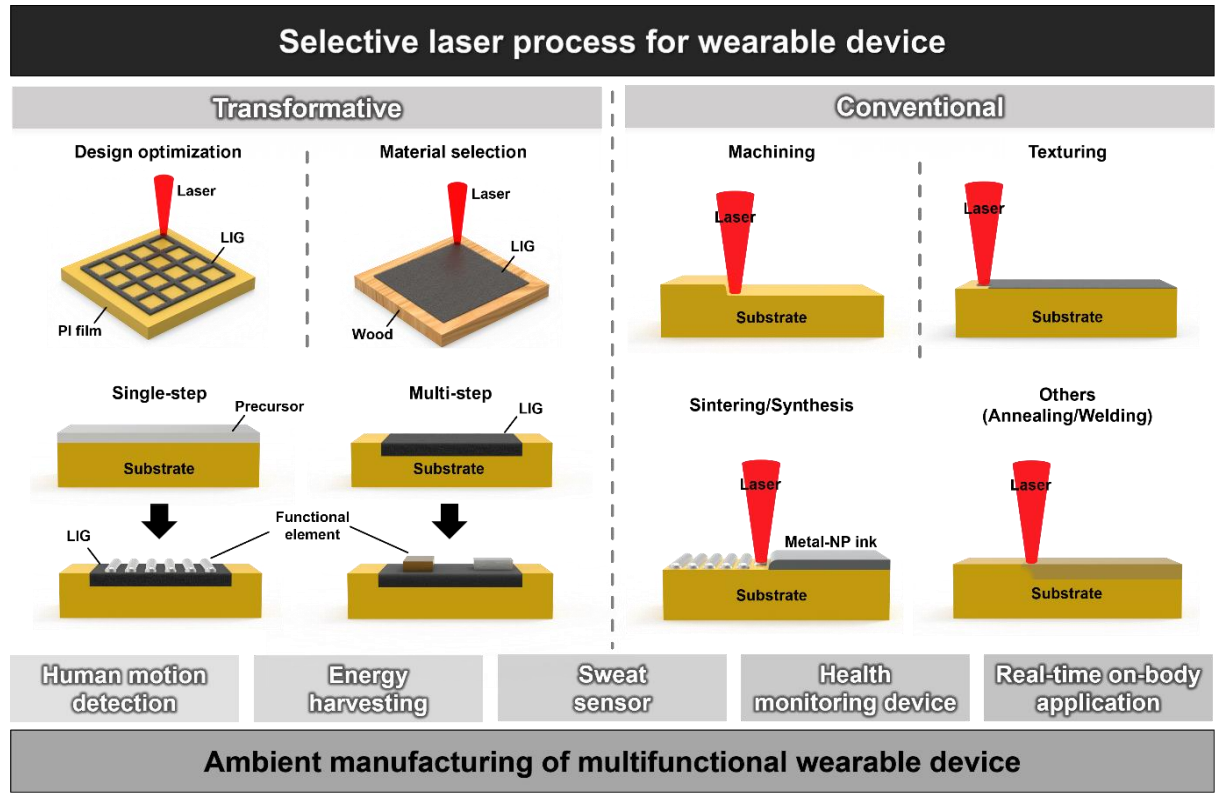

**Figure 1.** Overall classification of the selective laser processes for wearable devices that are divided into transformative and conventional laser processing. The proposed fabrication methods result in a multifunctional wearable device including human motion detection, energy harvesting, sweat diagnostics, health monitoring, and real-time on-body operation.

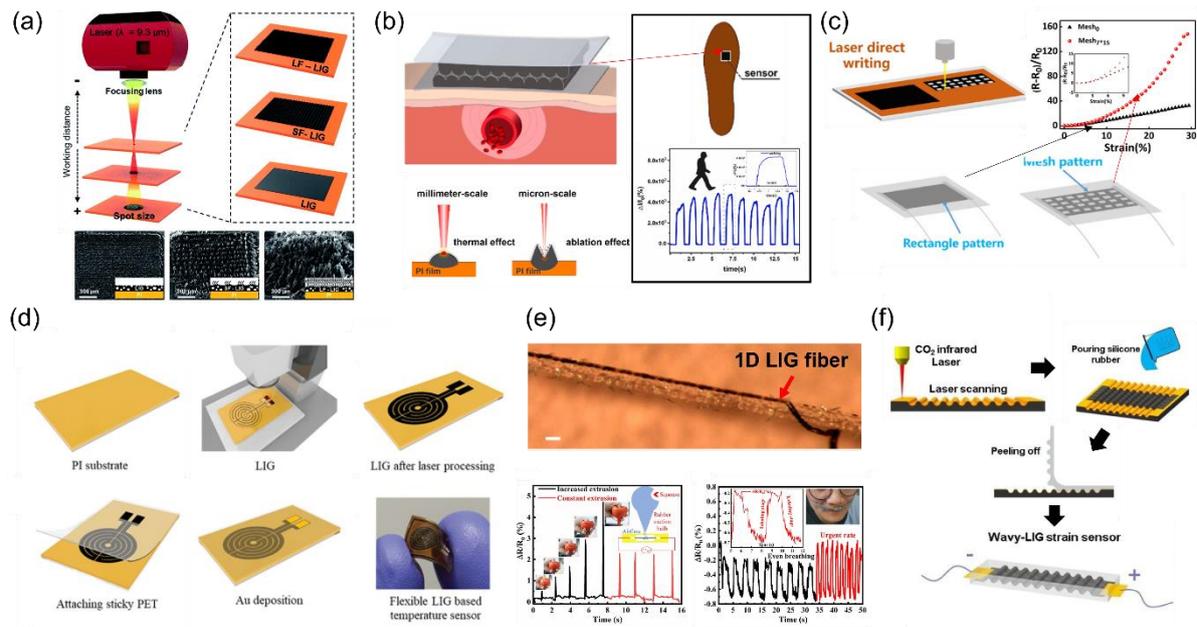

**Figure 2.** (a) Schematic illustration of LIG shape change according to working distance adjustment. SEM images of three different forms of LIG. Reprinted with permission from Ref.[9]. Copyrights 2020 Royal Society of Chemistry; (b) (Top) Schematic illustration of wave-shaped array LIG-based pressure sensor. (Bottom) Schematic diagram of morphologies of LIGs depending on laser spot size. (Left) Pressure sensor power change graph according to human movement. Reprinted with permission from Ref.[10]. Copyrights 2023 Elsevier; (c) Relative resistance change graph of strain sensor through LIG mesh patterning. Reprinted with permission from Ref.[12]. Copyright 2020 American Chemical Society; (d) Process schematic diagram of a circularly patterned LIG-based temperature detection sensor. Reprinted with permission from Ref.[13]. Copyrights 2020 Wiley-VCH Verlag GmbH & Co. KGaA, Weinheim; (e) Image of 1D LIG fiber. Performance graph of airflow sensing and respiratory monitoring. Reprinted with permission from Ref.[16]. Copyrights 2020 Elsevier; (f) Process schematic diagram of wavy-LIG strain sensor. Reprinted with permission from Ref.[14]. Copyrights 2020 MDPI.

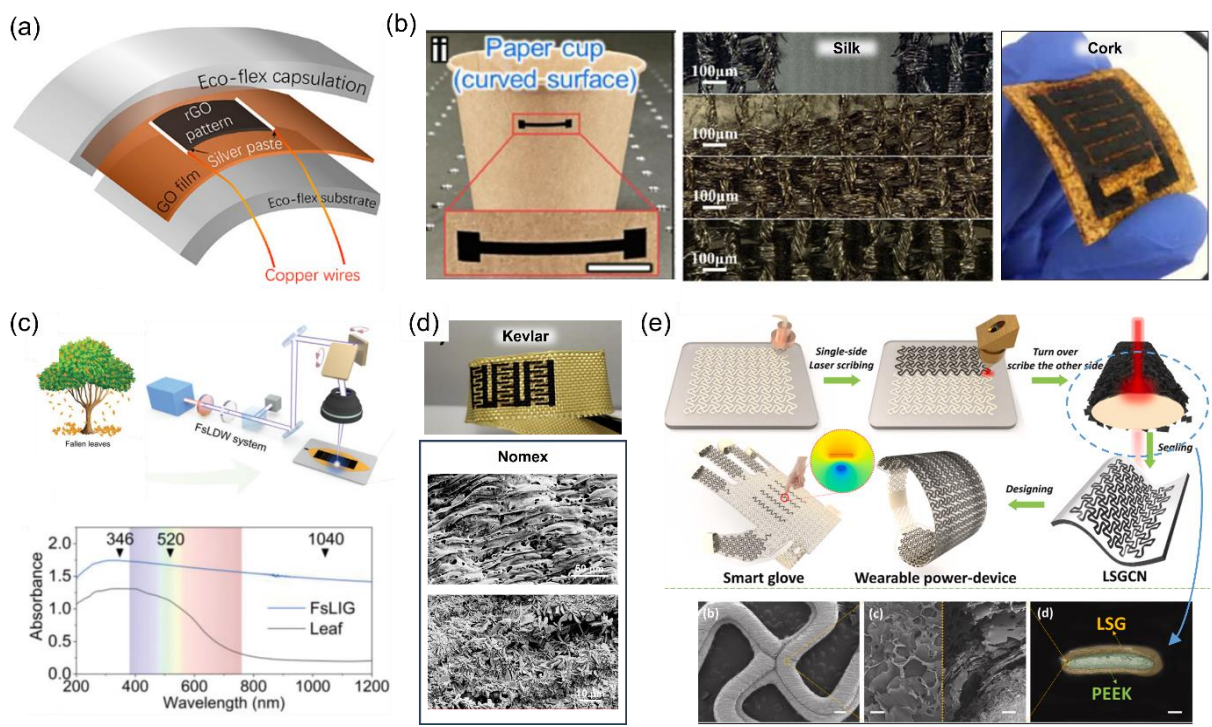

**Figure 3.** (a) Schematic layout of an LSG-based single sensor. Reprinted with permission from Ref.[27]. Copyright 2020 American Chemical Society; (b) (Left) Image of LIG patterned on paper. Reprinted with permission from Ref.[32]. Copyrights 2022 Elsevier; (Middle) Image of LIG patterned on silk. Reprinted with permission from Ref.[36]. Copyrights 2021 Wiley-VCH Verlag GmbH & Co. KGaA, Weinheim; (Right) Image of LIG patterned on cork. Reprinted with permission from Ref.[33] Copyrights 2022 IOP Publishing Ltd; (c) Process schematic illustration of the production of LIG with leaves. (Bottom) The light absorption spectrum of the leaf. Reprinted with permission from Ref.[37]. Copyrights 2021 Wiley-VCH Verlag GmbH & Co. KGaA, Weinheim; (d) (Top) Image of LIG patterned on Kevlar. Reprinted with permission from Ref.[46]. Copyrights 2023 Elsevier; (Bottom) SEM image of LIG patterned on pristine and copper-coated Nomex sheet. Reprinted with permission from Ref.[47]. Copyright 2021 American Chemical Society; (e) Fabrication processes of the peek-based smart glove and its detailed SEM images. Reprinted with permission from Ref.[48]. Copyrights 2022 Elsevier;

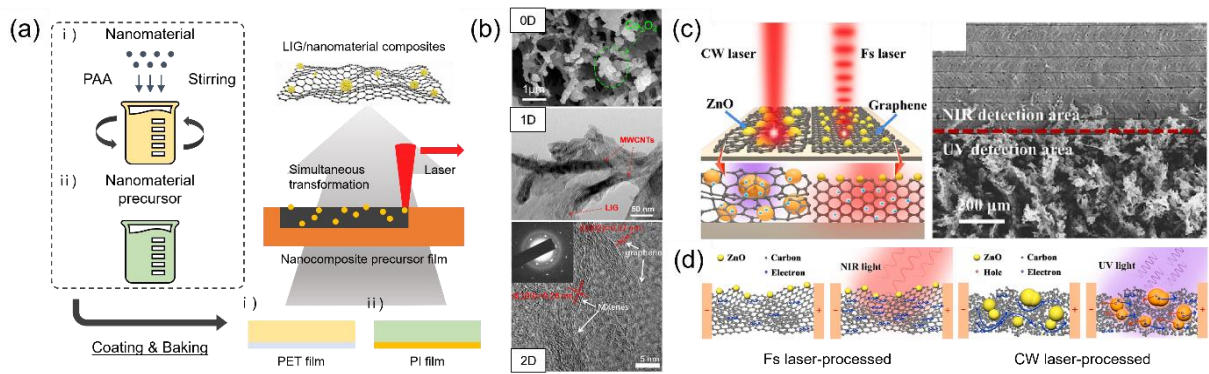

**Figure 4.** (a) Schematics of processing steps for simultaneous transformation to obtain LIG/nanomaterial composites. Reprinted with permission from Ref. [53]. Copyrights 2022 Elsevier; (b) SEM and TEM images of 0D, 1D, and 2D LDM integrated into the 3D LIG structure. Reprinted with permission from Ref. [62, 63, 64]. Copyrights 2022 Elsevier, Copyrights 2021 Elsevier, Copyrights 2023 American Chemical Society; (c) Schematic diagram and SEM image illustrating the difference between CW laser processing and Fs laser processing. (d) NIR light and UV light sensing mechanism of fs laser-processed surface and CW laser-processed surface. Reprinted with permission from Ref. [71]. Copyrights 2023 Elsevier;

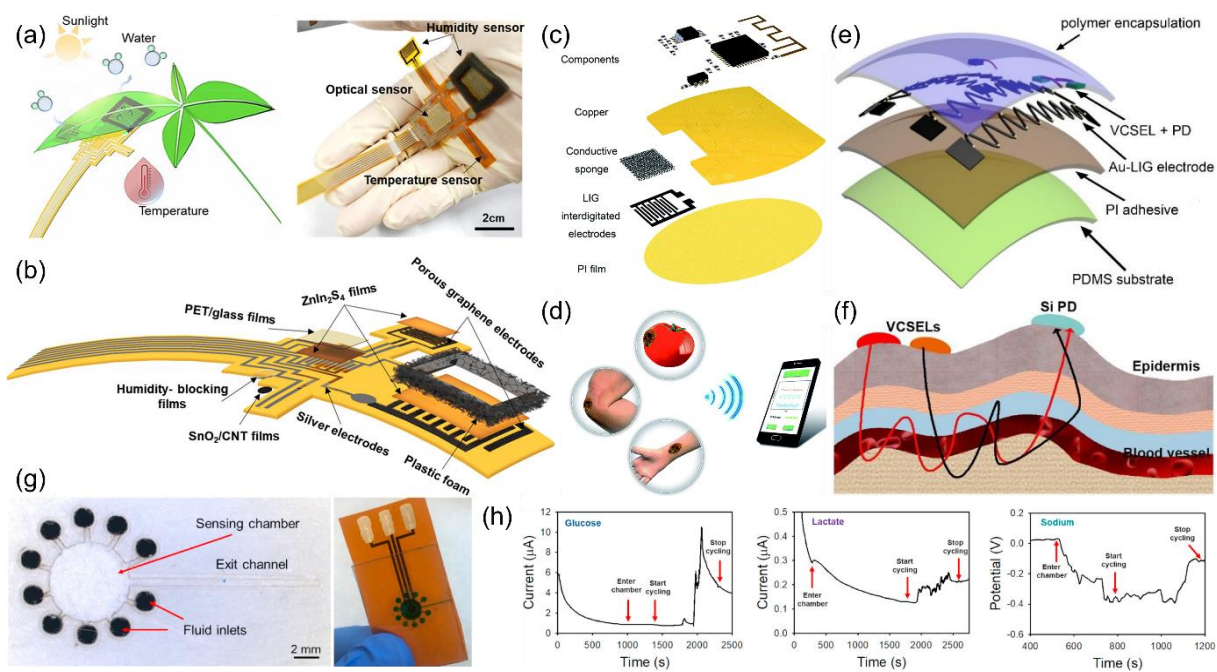

**Figure 5.** (a) Schematics and digital image of a flexible plant growth monitoring system. (b) Detailed configuration of the multimodal plant healthcare flexible sensor system. Reprinted with permission from Ref. [96]. Copyrights 2020 American Chemical Society; (c) Multilayered integration of the attachable strain sensing system. (d) Multifunctional applications of the sensor chip and the wireless communication. Reprinted with permission from Ref. [97]. Copyrights 2021 Royal Society of Chemistry; (e) Complete configuration of the skin-interfaced optoelectronic device. (f) Schematic of the blood oxygen sensing mechanism. Reprinted with permission from Ref. [98]. Copyrights 2022 American Chemical Society; (g) Digital images showing the microfluidic chamber design and its integration with patterned LIG sweat sensor on PI film. (h) Real-time on-body monitoring of the glucose, lactate, and sodium during cycling. Reprinted with permission from Ref. [99]. Copyrights 2023 American Chemical Society;

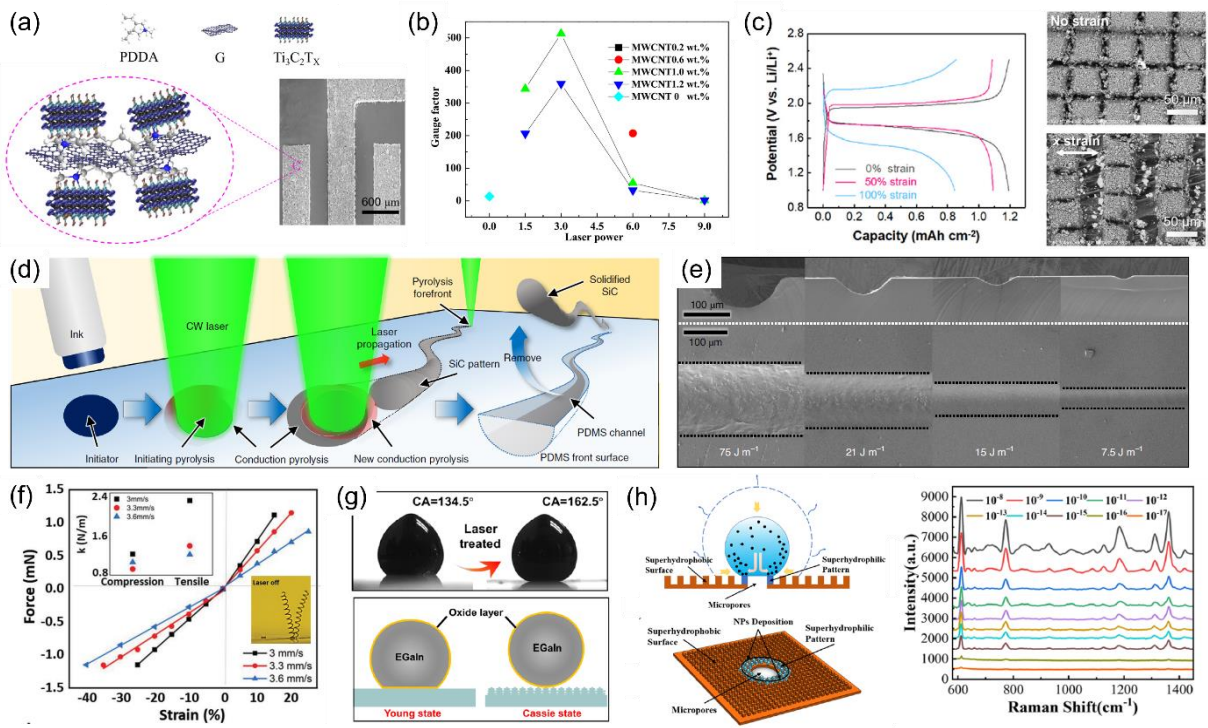

**Figure 6**. (a) Clean ablation feature of PDDA modified reduced graphene oxide with exfoliated Ti3C2Tx. Reprinted with permission from Ref. [102]. Copyrights 2022 Elsevier; (b) Gauge factor of MWCNT/PDMS having different MWCNT content processed at various laser power. Reprinted with permission from Ref. [104]. Copyrights 2020 Elsevier; (c) Discharge/charge voltage profiles of the laser-processed stretchable battery at different stretched states. (inset) SEM images of the stretchable LTO electrodes at different strain. Reprinted with permission from Ref. [109]. Copyrights 2022 Elsevier; (d) Schematic illustration of successive laser pyrolysis (SLP) process for PDMS micromachining. (e) Cross-sectional and top-view SEM images of PDMS after the application of SLP micromachining. Reprinted with permission from Ref. [117]. Copyrights 2021 Nature Publishing Group; (f) Force-strain surve of LIG spring created by pyrolytic jetting process. Reprinted with permission from Ref. [119]. Copyrights 2023 Wiley-VCH Verlag GmbH & Co. KGaA, Weinheim; (g) Contact angle of EGaIn droplet on PVA surface before and after laser texturing. Reprinted with permission from Ref. [127]. Copyrights 2020 American Chemical Society; (h) SERS sensor on laser-textured copper film. (Left) Schematics of evaporation process. (Right) Raman spectra of R6G at different concentrations. Reprinted with permission from Ref. [135]. Copyrights 2022 American Chemical Society.

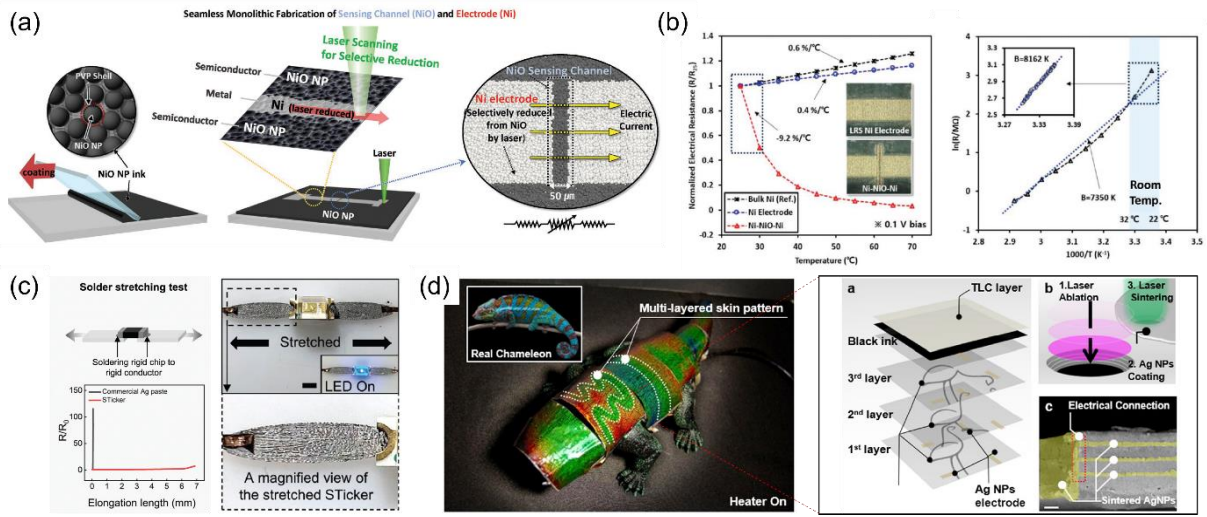

**Figure 7.** (a) Process illustration of the m-LRS process for monolithic Ni-NiO integration. (b) Temperature dependent electrical resistance showing NTC behaviour and unprecedentedly high B-value. Reprinted with permission from Ref. [142]. Copyrights 2023 Wiley-VCH Verlag GmbH & Co. KGaA, Weinheim; (c) Solder stretching test of STicker together with its optical photograph. Reprinted with permission from Ref. [145]. Copyrights 2023 Wiley-VCH Verlag GmbH & Co. KGaA, Weinheim; (d) Biomimetic chameleon soft robot. (Left) ATACS patch under operation. (Right) Configuration of the multilayered ATACS. Reprinted with permission from Ref. [146]. Copyrights 2021 Nature Publishing Group.

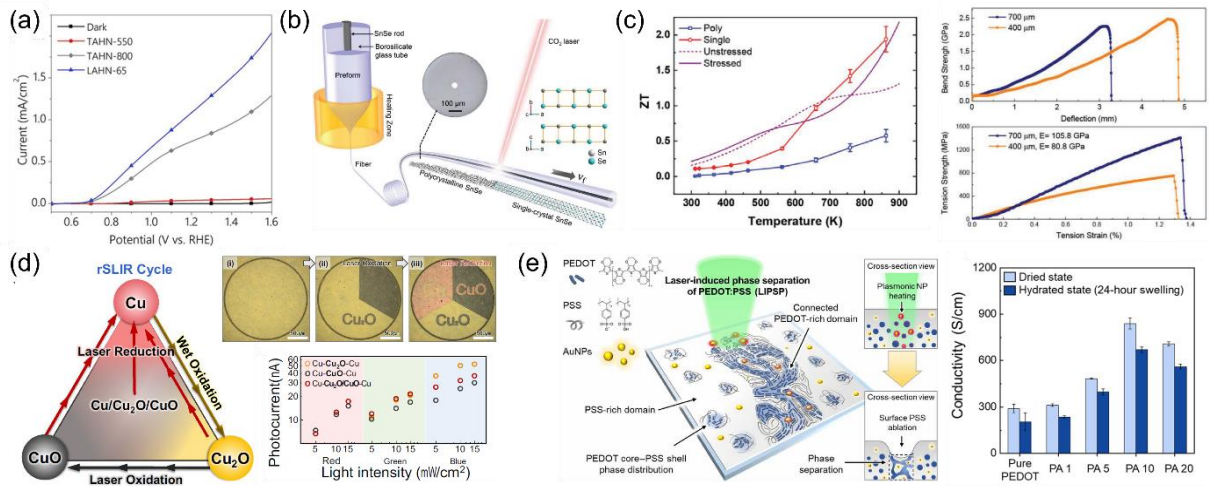

**Figure 8.** (a) Photocurrent density of PEC cells after thermal annealing and laser annealing. Reprinted with permission from Ref. [169]. Copyrights 2020 American Chemical Society; (b) Schematic illustration of thermal drawing and laser crystallization of SnSe fiber. (c) Thermoelectric figure of merit ZT value of poly- and single crystal SnSe fiber and their mechanical properties. Reprinted with permission from Ref. [157]. Copyrights 2020 Wiley-VCH Verlag GmbH & Co. KGaA, Weinheim; (d) Diagram of rSLIR cycle and photocurrent level for different laser-enabled material configuration. Reprinted with permission from Ref. [170]. Copyrights 2022 Springer-Verlag GmbH; (e) Schematic illustration of the laser-induced phase separation of PEDOT:PSS together with its electrical conductivity in the dried state and hydrated state. Reprinted with permission from Ref. [159]. Copyrights 2012 American Association for the Advancement of Science (AAAS).

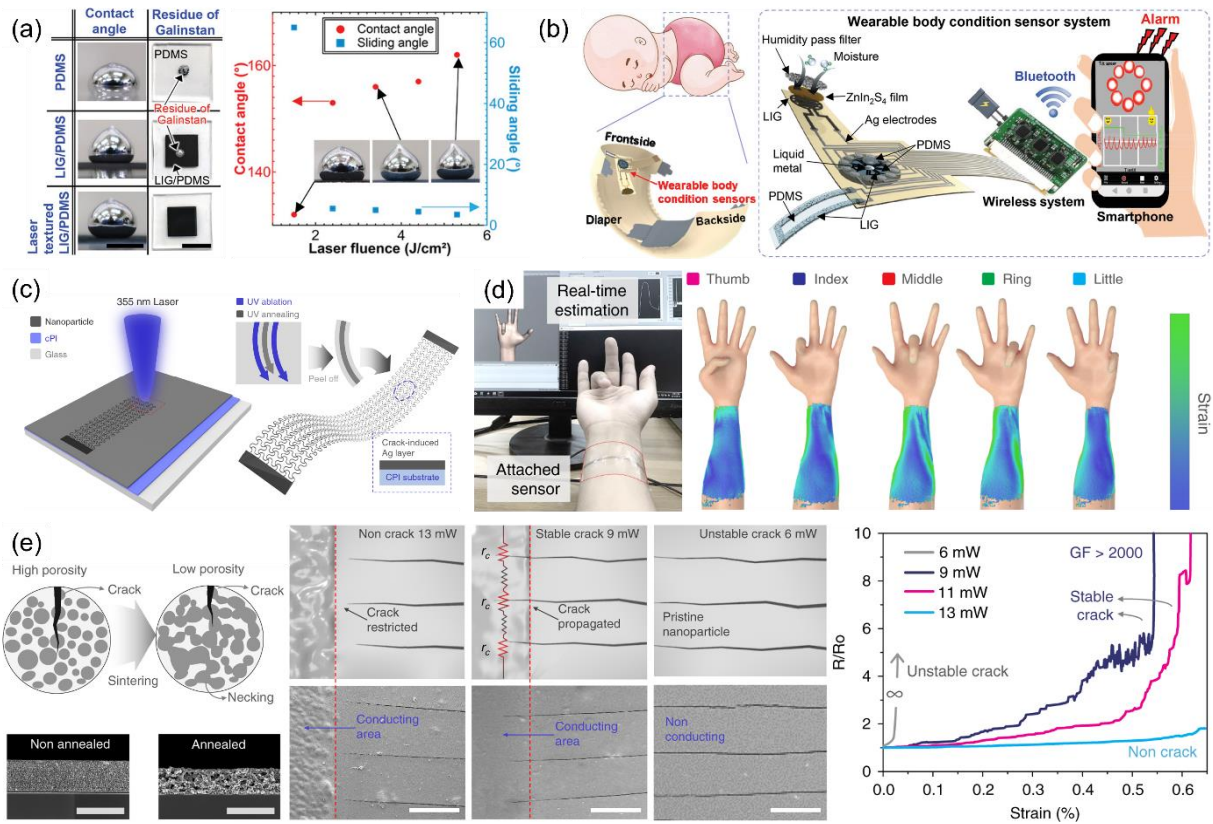

**Figure 9.** (a) Enhanced hydrophobicity of the laser textured LIG/PDMS surface. (b) Fully integrated systematic configuration of the wearable body condition sensor system. Reprinted with permission from Ref. [131]. Copyrights 2021 Wiley-VCH Verlag GmbH & Co. KGaA, Weinheim; (c) Schematics of the multistep laser processing. (d) Real-time estimation of the human finger motion by detection of the varying strains. (e) Programmable crack propagation depth by controlled laser fluence parameters and the resultant sensitivity of the laser-induced crack-based strain sensor. Reprinted with permission from Ref. [140]. Copyrights 2020 Springer Nature;

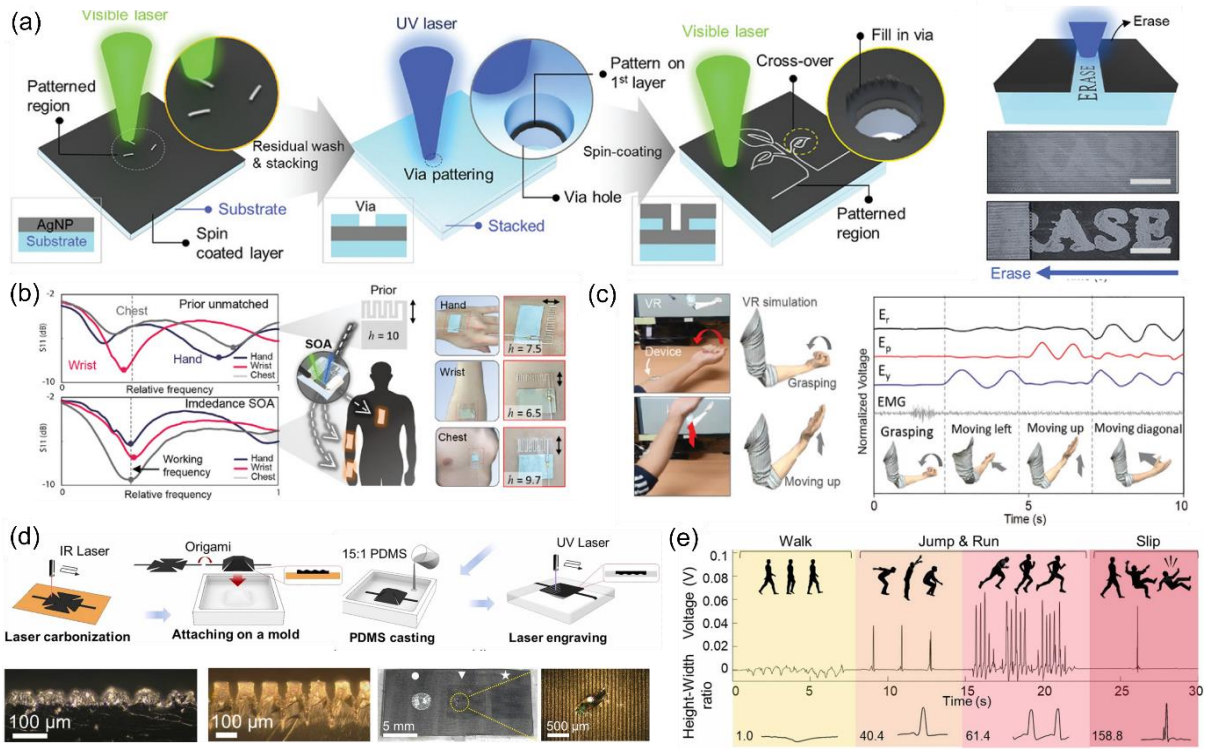

**Figure 10.** (a) Process schematic illustration of laser sintering and ablation processes using multiple lasers. . Reprinted with permission from Ref.[141]. Copyrights 2021 Wiley-VCH Verlag GmbH & Co. KGaA, Weinheim; (b) Different impedance optimization depending on skin attachment location. (c) Motion detection testing with wireless wearable sensors. (d) Fabrications schematic illustration of SSVS. (Bottom) Image of LIG/PDMS according to laser texturing. Reprinted with permission from Ref.[165]. Copyrights 2022 Elsevier; (e) Contextual motion detection graph using SSVS.

### <Table 1. Chemical sensor>

| Application | Laser process | Laser source | Materials | Performance | Ref |
|---|---|---|---|---|---|
| Sweat sensor | Laser-induced pyrolysis (LIP) | 450 nm laser | Paper | Sensitivity: 0.08 µA·µM$^{-1}$·cm$^{-2}$; LOD: 0.14 µM | [75] |
| | LIP | CO2 laser | PI film | Sensitivity: 2040 µA·mM$^{-1}$·cm$^{-2}$; LOD: 0.29 µM; Linear range: 0.5 ~1666 µM | [94] |
| | LIP | CO2 laser | Metal + LIG composite | Linear range: 1 ~100 µM; LOD: 0.47 µM | [95] |
| | LIP & Laser cutting | CO2 laser | PI film | Sensitivity: 2.47 × 10$^{-3}$; LOD: 220 µM; Linear range: 0.22 ~28 mM | [99] |
| | LIP & Laser ablation | CO2 laser | PI film + multilayer | Sensitivity: (UA) 3.50 µA·µM$^{-1}$ cm$^{-2}$; (Tyr)0.61 µA·µM$^{-1}$·cm$^{-2}$ | [160] |
| | Laser engraving | 800 nm fs laser | PEDOT:PSS + Carbon | Sensitivity: 0.875 µA·uM$^{-1}$·cm$^{-2}$; LOD: 1.2 µM; Linear range: 2 ~250 µM·L$^{-1}$ | [115] |
| | Laser engraving | CO2 laser | Ti3C2Tx MXene + LIG | LOD: 88 pM; Linear range: 0.01 ~100 nM | [171] |
| Glucose sensor | LIP | CO2 laser | PI film + Pt NPs coating + chitosan-glucose oxidase | Sensitivity: 4.622 µA·mM$^{-1}$ | [91] |
| | LIP | CO2 laser | Nomex + Cu plated | Sensitivity: (CuO-U electrode) 0.25 mA·cm$^{-2}$·mM$^{-1}$; (Cu-MS electrode) 0.32 mA·cm$^{-2}$·mM$^{-1}$ | [47] |
| | LIP | CO2 laser | PI film + Ni + Au | Sensitivity: 3500 µA·mM$^{-1}$·cm$^{-2}$ | [92] |
| | LIP | CO2 laser | PI film + PEDOT | Sensitivity: (glucose) 3.87 µA·mM$^{-1}$, (lactate) 11.83 µA·mM$^{-1}$ | [88] |
| | LIP | NA | PI film + Cu NPs | Sensitivity: 495 µA·mM$^{-1}$·cm$^{-2}$ | [90] |
| Gas sensor | LIP | CO2 laser | PI film + MoS2 casting | (NO$_2$ gas): - 47.6%; Response time: 2.87(m); cc(10 ppm) | [172] |
| | LIP | CO2 laser | PI film + elastomeric materials + Ag coating | (NO$_2$ gas): 5%; Response time: 360(s); Recovery time 720(s); LOD: 1.2ppb | [85] |
| | LIP | CO2 laser | PI film | (NO$_2$ gas): 6.66‰ ppm$^{-1}$; LOD: 4 ppb (NO gas): 4.18‰ ppm$^{-1}$; LOD: 8.3 ppb | [86] |
| | LIP | 532 nm (CW) | Paper | (NO$_2$ gas): 0.16% (40 ppm); LOD: 5 ppm | [32] |
| | LIG jetting | 532 nm (CW) | PI film | (NO$_2$ gas): 1‰ (5 ppm); LOD: 28 ppb | [119] |
| | Laser reduction | CO2 laser | Graphene + Mxene | (NO$_2$/NH$_3$ gas):1% ppm$^{-1}$; LOD: 5 ppb | [64] |
| Humidity sensor | LIP | CO2 laser | PI film + GO drop casting | 11%RH ~97%RH | [83] |
| | LIP | CO2 laser | LIG + ZIS | 40%RH ~90%RH | [96] |
| | LIP | CO2 laser | LIG + ZIS | 30%RH ~90%RH | [87] |
| | LIP | CO2 laser | Pb/HNb3O8 stacked on LIG electrodes | 30%RH ~90%RH | [84] |
| | SLS | 355 nm (ns) | Ga2O3 + LM on PI film | 30%RH ~95%RH | [143] |
| | Laser printing | 800 nm fs laser | PEDOT:PSS | 11%RH ~69%RH | [173] |
| | Laser ablation | CO2 laser | Cotton denim fabric + Conductive textile | 10%RH ~90%RH | [123] |
| | Laser annealing | 808 nm laser | Metal-oxide material (indium zinc oxide) | 10%RH ~60%RH | [151] |

<Table 2. Physical sensor>

| Application | Laser process | Laser source | Materials | Performance | Ref |
|---|---|---|---|---|---|
| **Strain sensor** | LIP | $CO_2$ laser | PI film + ZnSe | Gauge factor (GF): 107,428 (at -0.09 ~ 0.09% strain) | [20] |
| | LIP | $CO_2$ laser | PI film + Co NPC | GF: 1177 (0 ~18%), 39548 (18 ~23%) | [81] |
| | LIP | $CO_2$ laser | PI film + PDMS | GF: 37.8 (at 31.8% strain) | [14] |
| | LIP | $CO_2$ laser | PEEK | GF: 1.06 (0 ~60%) | [49] |
| | LIP | $CO_2$ laser | PI film + PDMS | GF: 3.54 (at 0 ~100% strain) | [17] |
| | LIP | $CO_2$ laser | PI film + Ecoflex with crack | GF: 114.8 (~50%) | [11] |
| | LIP | $CO_2$ laser | PI film + PDMS | GF: 950 (at 20 ~30% strain) | [12] |
| | LIP | 355 nm ps laser | PI fabric | GF: 27 (at 4% strain) | [15] |
| | LIP | 355 nm pulsed laser | PDMS + PSPI | GF: 21 ~35 (at strain range 0 ~1%) | [73] |
| | LIP | - | PDMS + PI particles | GF: linear range (3 ~79%), 59 (0 ~3%), 606 (3 ~7%), 1948 (7% ~) | [74] |
| | Laser-induced graphitization | 460 nm laser | Lignin paper | GF: (Tensile) 201 (strain: 0 ~0.025%), 255 (0.025 ~0.125%), 408 (0.125 ~0.225%); (Compressive) 91 (0 ~0.025%), 70 (0.125 ~0.225%) | [43] |
| | LIP & SLS | $CO_2$ laser | PI film + PDMS | GF: 243.42 (0 ~30%), 2710.95 (30 ~40%) | [62] |
| | LDW | 355 nm ps laser | PI film + Pt precursor + PDMS | GF: 45.6 (0 ~6%), 269.5 (6 ~17%), 489.3 (17 ~20%) | [53] |
| | rGO | 780 nm fs laser | GO +Ag | GF: 52.5 (in strain range of 25.4%) | [26] |
| | Laser cutting | 1060nm pulsed laser | Ti3C2–MXene | GF: 7400 (at 0.7% strain) | [101] |
| | Laser engraving | UV laser marker | Carbon-added PDMS + Ag | GF: tunable from 3.4 to 4570.6 | [110] |
| | Laser ablation | $CO_2$ laser | PDMS/MWCNT composite | GF: 513.2 (at 5% strain) | [104] |
| | Laser ablation & Laser patterning | 515 nm fs laser | PDMS + Graphene | GF: 496.7 | [106] |
| | Laser ablation & Laser annealing | 355 nm ns laser | Crack-induced Ag layer on cPI | GF: >2000 | [140] |
| **Pressure sensor** | LIP | $CO_2$ laser | PI film + PU transfer | Sensitivity: 149 kPa$^{-1}$ (0 ~1 kPa), 659 kPa$^{-1}$ (1 ~10 kPa), 2048 kPa$^{-1}$ (10 ~100 kPa) | [18] |
| | LIP | $CO_2$ laser | PI film + Ag NP on MX sponge | Sensitivity: 0.9 kPa$^{-1}$ at 0.5 kPa | [97] |
| | LIP | $CO_2$ laser | PI film + LIG powder | Sensitivity: 1.86 kPa$^{-1}$ at 0 ~150 Pa | [19] |
| | LIP | $CO_2$ laser | MWCNTs + PDMS + PI film | Sensitivity: 2.41 kPa$^{-1}$ at 0 ~200 Pa | [63] |
| | LIP | $CO_2$ laser | PI film + PDMS | Sensitivity: 85 × 10$^{-4}$ kPa$^{-1}$ at 0 ~30 kPa, 12 × 10$^{-4}$ kPa$^{-1}$ at 30 ~220 kPa | [17] |
| | LIP | $CO_2$ laser | PI film + Ecoflex with crack | Sensitivity: 1.64 × 10$^{-2}$ kPa$^{-1}$ at 0 ~120 kPa | [11] |
| | LIG | $CO_2$ laser | PI film + GO cloth | Sensitivity: 30.3 kPa$^{-1}$ at 0 ~2.5 kPa, 0.56 kPa$^{-1}$ at 2.5 ~20 kPa | [80] |
| | LDW | 915 nm laser | PI film | Sensitivity: Linear range at 0 ~100 kPa; Resolution: mPa scale | [10] |
| | rGO | $CO_2$ laser | GO + Cellulose fiber | Sensitivity: 19.47 kPa$^{-1}$ at 0 ~4 kPa | [42] |
| | LSG | 450nm laser | Graphene oxide | Sensitivity: 434 kPa$^{-1}$ at 200 kPa | [27] |
| | Laser ablation | 355 nm fs laser | PDMS + Ag NW coating | Sensitivity: 4.48 kPa$^{-1}$ at 0 ~22 kPa | [113] |
| | Laser ablation | 800 nm fs laser | PDMS + LM | Sensitivity: ~2.78 kPa$^{-1}$ | [129] |
| **Temperature sensor** | LIP | 532 nm (CW) | Paper | Temperature resistance coefficient, R(T): -0.15% ℃$^{-1}$ | [32] |
| | LIP | 1064 nm pulsed laser | PI film | R(T): 0.00142 °C$^{-1}$ with high linear ($R^2$ = 0.999) | [13] |
| | rGO | $CO_2$ laser | GO + Cellulose fiber | R(T): -0.195% ℃$^{-1}$ at 30 ~50 ℃ | [42] |
| | SLS | 532 nm (CW) | Ni NP | (Heater): 310 ℃ at 11 V | [174] |
| | Laser welding | 940 nm laser | Metal foil textile | R(T): 0.0039 °C$^{-1}$, $R^2$ of 0.997 | [154] |

| | Laser ablation | 355 nm ns laser | Thermoelectric pellet | (Thermo-haptic device): Cooling: 11 ℃; Heating: 40 ℃ | [120] |
|---|---|---|---|---|---|
| | Laser curing & Laser patterning | 1064 nm pulsed laser | Si + Polymer | (Heater): 80 °C at 3 W·cm$^{-2}$ | [152] |
| | Laser-induced reductive sintering | 532 nm (CW) | Ni-NiO-Ni | R(T): -9.2% ℃$^{-1}$ (20 ~30 ℃), B-value: 7350 K (25-70 ℃), 8162 K (25-30 ℃); | [142] |
| **Optical sensor** | LIP & SLS | CO2 (CW) & 800 nm fs laser | ZnO NP on PI film | (Photodetector) Responsivity: 0.1 ~0.9 mA·W$^{-1}$ (2.4 ~3.1 W); Dual-mode (UV/IR) | [71] |
| | LIHG | 532 nm (CW) | ZnO NW + CuO NW | (Photodetector) Responsivity: 2.39 × 10$^{-4}$ A·W$^{-1}$ (bias: 2 V); Photocurrent density: 1.25 A·cm$^{-2}$; | [148] |
| | Laser-induced redox | 532 nm (CW) | Cu | (Photodetector) Multispectral to RGB | [170] |
| | Laser texturing | 520 nm fs laser | AISI304 stainless steel | (SERS sensor) LOD: 10$^{-14}$ M | [134] |
| | Laser drilling & Laser ablation & Laser texturing | 520 nm fs laser | NA | (SERS sensor) LOD: 10$^{-17}$ M | [135] |
| | Laser carbonization & Graphitization | 400~460 nm laser | resorcinol-formaldehyde (RF) aerogel | (Photodetector) Sensitivity: 0.12%·mW$^{-1}$ at infrared light | [41] |
| | SiC sintering? | 800 nm fs laser | SiC NPs | (Photodetector) Responsivity: 55.89 A·W$^{-1}$ (bias:1 V) at UV light | [149] |
| **Sound sensor** | LIP | CO2 laser | PI film | Sound pressure level: 49dB at 20 kHz | [175] |
| | Laser ablation | NA | Kapton + PI + PVDF + Ag | Resonant frequency at 199.37kHz | [112] |
| | LIP & Laser engraving | CO2 & 355 nm (ns) | LIG + PDMS | Peak acc: 6 ~68.5 m·s$^{-2}$ | [165] |

<Table 3. Energy device>

| Application | Laser process | Laser source | Materials | Performance | Ref |
|---|---|---|---|---|---|
| **Triboelectric nanogenerators** | LIP | $CO_2$ laser | PI film | Power density: 512 mW·m$^{-2}$ | [9] |
| | LIP | $CO_2$ laser | PI film + PTEE coating/GO-Cu coating | Power density: 41 μW·cm$^{-2}$ | [79] |
| | LIP | $CO_2$ laser | PEEK | Power density: 2.3 mW·cm$^{-2}$ | [49] |
| | LIP | $CO_2$ laser | PI film + rGO-cloth | Transfer charge density: 270 μC·m$^{-2}$ | [80] |
| | Laser ablation | $CO_2$ laser | FPCB | Power density: 416 mW·m$^{-2}$ | [176] |
| | Laser ablation | 355 nm pulsed laser | PI film + SRPA + Al | Power density: 98.35 mW·m$^{-2}$ | [111] |
| | Laser synthesis | 1064 nm pulsed laser | Si + $MoS_2$ | Power: 2.25 μW | [28] |
| | Laser reduction | $CO_2$ laser | Graphene/MXene | Power density: 57 mW·cm$^{-2}$ | [64] |
| **Thermoelectric generator** | Laser ablation | 1064 nm pulsed laser | TE film | Power density: 1.04 mW·cm$^{-2}$ | [108] |
| | Laser engraving | $CO_2$ laser | Ag + Ti + PDMS films | Power: 7.9 μW | [122] |
| **Moisture-driven power generator** | LIP | $CO_2$ laser | Cellulose nanofiber | Power density: 4.92 μW·m$^{-2}$ | [39] |
| **Supercapacitor** | LIP | $CO_2$ laser | PI film + $Co_3O_4$ precursor | Capacitance: 10.9 mF·cm$^{-2}$ at 5 mV·s$^{-1}$; $PVA/H_2SO_4$ | [54] |
| | LIP | $CO_2$ laser | PI film + KOH | Capacitance: 244 μF·cm$^{-2}$ at 100 mV·s$^{-1}$; PVA/H$^+$ | [50] |
| | LIP | $CO_2$ laser | PI film + KOH | Capacitance: 32 mF·cm$^{-2}$ at 20 mV·s$^{-1}$; $PVA/H_3PO_4$ | [51] |
| | LIP | 450 nm laser | PI film | Capacitance: 8.11 mF·cm$^{-2}$ at 100 mV·s$^{-1}$; $PVA/H_2SO_4$ | [21] |
| | LIP | $CO_2$ laser | Carbon cloth + $MoO_2$ | Capacitance: 81.8 mF·cm$^{-2}$ at 10 mV s$^{-1}$; Mo ion ink | [55] |
| | LIP | $CO_2$ laser | PAA film + $MoS_2$ | Capacitance: 35.3 mF·cm$^{-2}$ at 5 mV s$^{-1}$; PVA/NaOH | [52] |
| | LIP | $CO_2$ laser | PI film + PTFE + Phosphor copper sheets | Capacitance: 389.6 μF·cm$^{-2}$ at 10 mV·s$^{-1}$; $PVA/Na_2SO_4$ | [77] |
| | LIP | 346 nm fs laser | Leaf | Capacitance: 34.68 mF·cm$^{-2}$ at 5 mV·s$^{-1}$; $PVA/H_2SO_4$ | [37] |
| | LIP | 1064 nm laser | Cork | Capacitance: 1.35 mF·cm$^{-2}$ at 5 mV·s$^{-1}$; $PVA/H_2SO_4$ | [33] |
| | LIP | 450 nm laser | Paper | Capacitance: 166.6 mF·cm$^{-2}$ at 50 mV·s$^{-1}$; $H_2SO_4$ | [75] |
| | LIP | 532 nm fs laser | Lignin | Capacitance: 0.19 mF·cm$^{-2}$ at 1500 mV·s$^{-1}$; NaCl | [34] |
| | LIP | 450 nm laser | Lignin-based fiber membranes + $MoS_2$ | Capacitance: 527.8 F·g$^{-1}$ at 10 mV·s$^{-1}$; $PVA/H_2SO_4$ | [44] |
| | LIP | $CO_2$ laser | Kevlar textile | Capacitance: 125.35 mF·cm$^{-2}$ at 100 mV·s$^{-1}$; $PVA/H_2SO_4$ | [46] |
| | LIP | 532 nm laser | PI film | Capacitance: 1.59 mF·cm$^{-2}$ at 100 mV·s$^{-1}$; $Na_2SO_4$ | [119] |
| | LIP + Synthesis | $CO_2$ laser | PI film + Ni/Co ions ink/WPU transfer | Capacitance: 2.4 mF·cm$^{-2}$ at 10 mV·s$^{-1}$; PVA | [60] |
| | LIP + Synthesis | 532 nm pulsed laser | PEEK + $MnO_2$ coating | Capacitance: 48.9 mF·cm$^{-2}$ at 10 mV·s$^{-1}$; $PVA/H_2SO_4$ | [48] |
| | SLS | 355 nm pulsed laser | SWCNTs/MWCNTs + RGO/Ag-NWs | Capacitance: 4 F·cm$^{-2}$ at 5 mV·s$^{-1}$; $PVA/H_3PO_4$ | [137] |
| | SLS | 532 nm laser | PVDF + AgNPs | Capacitance: 24.5 mF·cm$^{-2}$ at 50 mV·s$^{-1}$; $PVA/Na_2SO_4$ | [138] |
| | Laser scribing | $CO_2$ laser | PI film + PEDOT | Capacitance: 115.2 F·g$^{-1}$ at 10 mV·s$^{-1}$; Aqueous electrolyte | [76] |
| | Laser reduction | 1030 nm fs laser | GO + PDMS | Capacitance: 52 F·cm$^{-3}$ at 5 mV·s$^{-1}$; $PVA/H_2SO_4$ | [23] |
| | Laser deposition | KrF excimer laser | $V_2O_5$ & $WO_3$ thin films | Capacitance: 40.28 F·cm$^{-3}$ at 500 mV·s$^{-1}$; PVA/KOH | [177] |
| | Laser engraving | $CO_2$ laser | PI film + $MnO_2$ | Capacitance: 15.04 mF·cm$^{-2}$ at 5 mV·s$^{-1}$; PVA/KOH | [61] |
| | LDW | UV laser | $Ti_3C_2Tx$ MXene | Capacitance: 241 mF·cm$^{-2}$ at 100 mV·s$^{-1}$; $PVA/H_2SO_4$ | [102] |
| | Laser ablation | 1064 nm laser | PEDOT:PSS-aramid nanofiber | Capacitance: 15.4 mF·cm$^{-2}$ at 2 mV·s$^{-1}$; $PVA/H_3PO_4$ | [103] |
| **Fuel cells** | LIP + ablation | 450 nm laser | Paper | Power density: 27 μW·cm$^{-2}$ | [75] |
| | Laser scribing | $CO_2$ laser | PI film/glucose dehydrogenase + bilirubin oxidase | Power density: 27 ± 1.7 μW·cm$^{-2}$ | [78] |
| **Battery** | LIP + SLS | $CO_2$ laser | PI film + $Co_3O_4$ precursor | Capacitance: 712 mAh·g$^{-1}$ | [62] |
| | Laser engraving | $CO_2$ laser | MWCNTs-$MnO_2$ or MWCNT-Zn | Capacitance: 116.6 μAh·cm$^{-2}$ | [105] |
| | Laser ablation | 1064 nm ps laser | $LiFePO_4$/$Li_4Ti_5O_{12}$ powders | Capacitance: 1.2 mAh·cm$^{-2}$ | [109] |